\documentclass[showpacs,aps,prl,twocolumn,superscriptaddress,10pt]{revtex4-2}
\usepackage[utf8]{inputenc}
\usepackage{graphicx} 
\usepackage{bm} 
\usepackage{color}
\usepackage{caption}
\usepackage{subcaption}
\usepackage{ulem}
\usepackage{soul}
\usepackage{amsmath}
\usepackage{amssymb}
\usepackage{enumerate}
\usepackage{xspace}
\usepackage{mathrsfs}
\usepackage{mathptmx}
\usepackage[colorlinks=true,linkcolor=blue,citecolor=blue,urlcolor=blue]{hyperref}
\usepackage{orcidlink}

\newcommand{\vampire}{\textsc{vampire} }
\newcommand{\be}{\begin{equation}}
\newcommand{\ee}{\end{equation}}
\newcommand{\dth}{\delta\theta_h}
\newcommand{\dts}{\delta\theta_s}
\begin{document}
%
\title{Higher-order magnetic anisotropy in soft-hard magnetic materials}
\author{Nguyen Thanh Binh \orcidlink{0000-0001-5869-6088}}
\affiliation{Department of Physics, University of York, York, YO10 5DD, UK}
\thanks{Corresponding author: btn500@york.ac.uk}
\author{Sarah~Jenkins\orcidlink{0000-0002-6469-9928}}
\affiliation{Department of Physics, University of York, York, YO10 5DD, UK}
\affiliation{TWIST Group, Institut f\"ur Physik, Johannes Gutenberg-Universit\"at, 55128 Mainz}
\affiliation{TWIST Group, Institut f\"ur Physik, University of Duisburg-Essen}
%
\author{Sergiu Ruta\orcidlink{0000-0001-8665-6817}}
\affiliation{Department of Physics, University of York, York, YO10 5DD, UK}
\affiliation{The Department of Engineering and Mathematics, Sheffield Hallam University, Sheffield, S1 1WB, UK}
\author{Richard~F.~L.~Evans\orcidlink{0000-0002-2378-8203}}
\affiliation{Department of Physics, University of York, York, YO10 5DD, UK}
\author{Roy~W.~Chantrell\orcidlink{0000-0001-5410-5615}}
\affiliation{Department of Physics, University of York, York, YO10 5DD, UK}

\begin{abstract}
We have computationally studied the properties of higher-order magnetic anisotropy constants in an $L1_0$/$A$1-FePt coreshell system which is characterized by a strong second-order 2-ion Fe-Pt anisotropy component. We show that the coreshell structure induces an unexpected fourth-order anisotropy constant $K_2$ the magnitude of which varies non-monotonically with the core-size ratio $R$ reaching a peak at $R \approx 0.50$. Furthermore, we find that $K_2$ scales with the normalized magnetization by $(M/M_s)^{2.2}$ at temperatures below the Curie temperature - a remarkable deviation from the established Callen-Callen theory which instead predicts a scaling exponent of 10. We construct an analytic model which demonstrates $K_2$ arises from the canting of the core and shell magnetization, and successfully reproduces and justifies the scaling exponent obtained from numerical simulation.
\end{abstract}
%
%
\maketitle

\begin{figure}[!tb]
\centering
 \includegraphics[width=\linewidth]{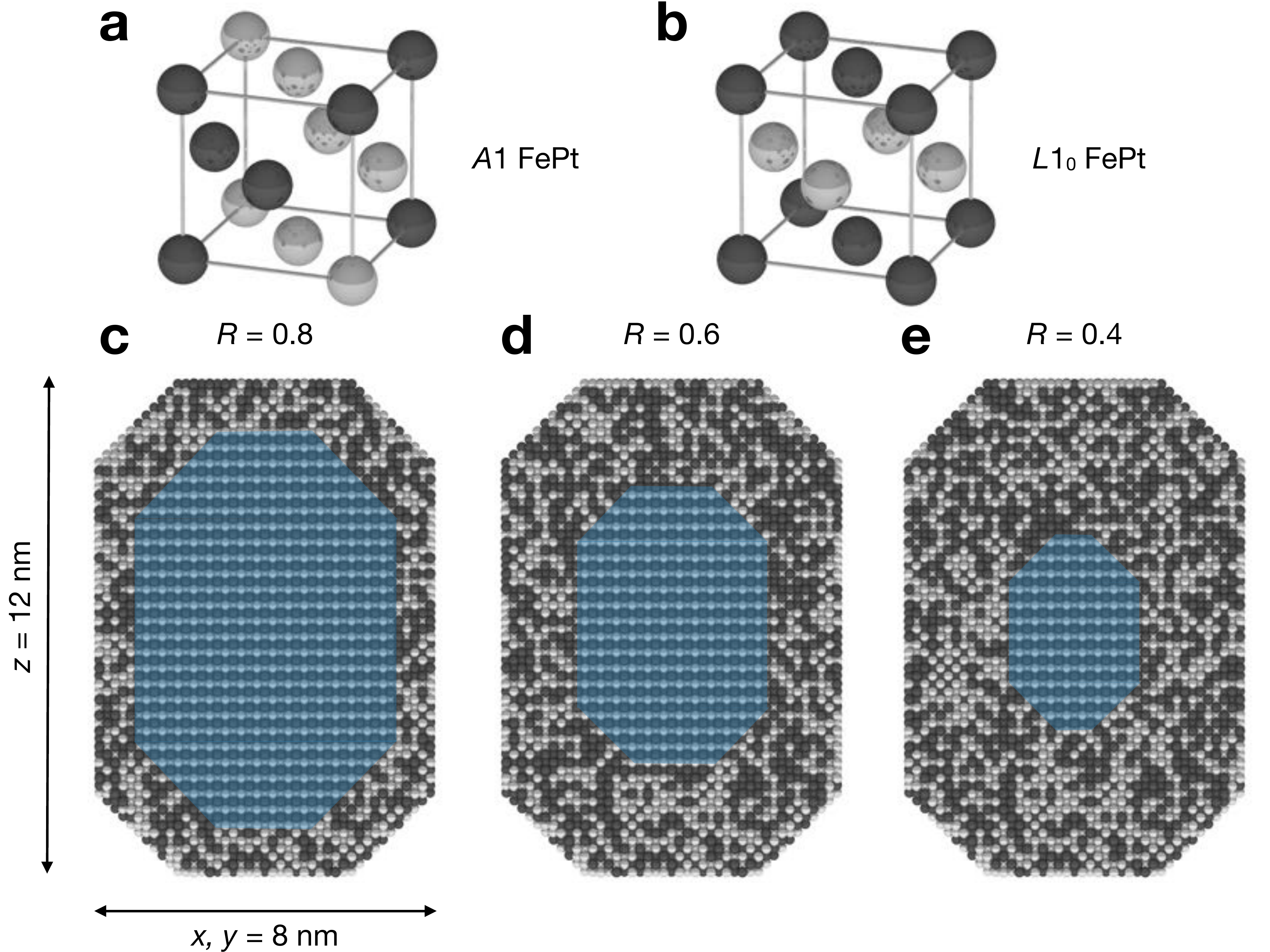}
 
   \caption[]{The crystal structures of FePt the disordered (a) A1-fcc bulk-alloy Fe$_{0.5}$Pt$_{0.5}$ and ordered (b) $L1_0$-fct. Dark and light spheres indicate Fe and Pt atoms respectively. Cross-sectional views of the simulated $L1_0/A1$ core-shell grains with core-size ratios $R = 0.80$ (c), $R = 0.60$ (d), and $R = 0.40$ (e) showing different volume fractions of the ordered and disordered phases. The core ordered region in each cross-section is indicated by the shaded area.}
\label{Fig_Coreshell}
\end{figure}

Heat-Assisted Magnetic Recording (HAMR) is emerging as the next-generation approach for magnetic recording~\cite{Rottmayer2006,Mcdaniel2005}. The functioning of HAMR requires the writing medium to be made of a magnetic material with high anisotropy and low Curie temperature. FePt in the $L1_0$ phase satisfies this requirement, and thus has been studied extensively for potential HAMR applications~\cite{Weller_HAMR,Kryder}. As prepared, bulk-alloy FePt generally exists in the A1 phase in which Fe and Pt atoms are randomly distributed, thus resulting in a low magnetic anisotropy. However, at high temperatures FePt can undergo a transition to the ordered $L1_0$ phase~\cite{Nakaya}, sketched in Fig.~\ref{Fig_Coreshell}. The exceptionally large magnetocrystalline anisotropy of $L1_0$-FePt stems from a hybridzation between the $3d$ Fe and $5d$ Pt orbitals along the [001] crystal direction~\cite{Daalderop,Sakuma,Oppeneer,Ravindran,Shick,Solovyev} which brings into play the strong spin orbit coupling of the Pt, resulting in a dominant 2-ion anisotropy component~\cite{Okamoto,Weller_Anisotropy} in addition to the local single-site anisotropy. 

Measurements of the second-order anisotropy constant $K_1$ in bulk $L1_0$-FePt using a simple angular form of the magnetic anisotropy energy function, $E=K_1\sin^2\theta$, have generally been consistent and well established~\cite{Okamoto,Bublat,Weller_Anisotropy,Shima,Inoue,Richter} with values for the magnetic anisotropy energy as high as 6.2 $MJ/m^3$~\cite{Richter}. On the contrary, there has not been a consensus on the existence and the significance of the fourth-order anisotropy constant $K_2$. Previous studies have arrived at conflicting conclusions where $K_2$ has been argued to be a misinterpretation~\cite{Richter}, negligibly small compared to $K_1$~\cite{Inoue}, or non-negligible~\cite{Okamoto}. In addition, a further issue drawing attention is a reported deviation of the scaling of $K_2$~\cite{Inoue} from the classical Callen-Callen power law~\cite{Callen1966TheLaw} which, interestingly, has also been observed in other materials~\cite{Chatterjee,Miura}.

Furthermore, a recent study by Sepehri-Amin \textit{et al.}~\cite{SepehriAmin} on $L1_0$-FePt thin films discovered an effect of Pt enrichment on the film surface regardless of the FePt composition, which subsequently was shown to distort the ordered structure of the $L1_0$ phase and thus reduce the magnetocrystalline anisotropy of the FePt grains. This phenomenon is manifested via a heavy intermixing of Fe atoms and Pt atoms on the grain surface. The proportions of Fe and Pt atoms at various distances from the grain surface are shown to be dependent on grain size. The intermixing between Fe and Pt atoms at the grain surface compromises the chemical ordering of $L1_0$-FePt grain, thus reducing the uniaxial anisotropy. The impact of this Pt segregation is found to be more pronounced in grains smaller than 15 nm, which is detrimental for HAMR where smaller grain sizes are desired. The varying relative proportion of the two $L1_0/A1$ phases has also been seen to affect the uniaxial magnetic anisotropy in previous experimental studies of phase-graded thin films~\cite{Barucca}. The effect of Pt surface segregation in a phase-graded FePt system, therefore, necessitates an investigation into quantifying the impact of $L1_0/A1$ phase composition to the anisotropy of $L1_0$-FePt.

In this letter we present a computational study using an atomistic model showing the existence of a fourth-order anisotropy component of phase-coupled $L1_0/A1$-FePt coreshell grains which are specifically constructed to replicate the aforementioned Pt surface segregation effect. We propose an analytic model to explain the properties of this fourth-order anisotropy and show that the applicability of our analytic model can be extended to a generic nanocomposite material with soft-hard magnetic interlayers.
%

We construct elongated FePt grains with faceted surfaces following the method of Moreno \textit{et al.}~\cite{Moreno2020} which closely resemble realistic ones found in a typical HAMR recording medium~\cite{SepehriAmin}. The grains are elongated along the [001] lattice direction by a shape factor of 1.5. In order to replicate the effect of Pt surface segregation, the grains are structured with a core made of the ordered $L1_0$-phase FePt surrounded by a disordered $A1$-phase FePt shell. The grain size is fixed at 8 nm $\times$ 8 nm $\times$ 12 nm. The diameter of the core can be freely adjusted so as to reproduce varying degrees of ordering through a surface coupling effect between the $L1_0$ and $A1$ phase. The fractional core size $R$ of the grain is defined as $R=d_{\mathrm{shell}}/d_{\mathrm{grain}}$ where $d_{\mathrm{core}}$ is the core diameter and $d_{\mathrm{grain}}$ the entire grain diameter. In simulations $R$ is varied between 0.05 and 0.95. The lower and upper bounds of $R$ represent two extreme cases: when $R=0.05$ the $L1_0$ core consists of only one single atom, while when $R=0.95$ the grain has only one atomistic layer of the $A1$ shell. Cross-sectional views of the core-shell grains with various core sizes $R$ are shown in Fig.~\ref{Fig_Coreshell}. The unit cell of the fct $L1_0$-FePt is slightly compressed on the c-axis~\cite{Alamgir,Lyubina,Klemmer} while that of the fcc $A1$-FePt is not. For simulation efficiency and without altering any physical properties, a common unit cell is implemented for both phases with a uniform cubic shape and a lattice spacing of $a=0.3795$ nm obtained from experiments and consistent with previous computational studies~\cite{Loc,Lyubina,Klemmer,Suttipan}.

Our simulations are carried out using the \vampire atomistic simulation software package~\cite{vampireURL,vampire} using a constrained Monte-Carlo (CMC) integrator~\cite{AsselinPRB2010}. The system magnetization can be constrained at an angle $\theta$ to the easy axis which is oriented along the $z$-direction. At temperatures varying from 0K to 1000K, a full angular sweep is performed for $\theta$ from 0 to 180 degrees. The anisotropy constants are computed via the angular-dependent restoring torque $\partial E(\theta)/ \partial \theta$ ~\cite{AsselinPRB2010,Wang_TorqueMethod,Wu_TorqueMethod}. For a uniaxial system, $E(\theta)$ can be expressed as a power series: $E = E_0 + K_1\sin^2(\theta) + K_2\sin^4(\theta) + ...$, where the constant $E_0$ can usually be omitted and $K_1$ and $K_2$ are the second and fourth-order anisotropy constants respectively. Therefore, by fitting to the torque computed from simulation output, the values of anisotropy constant(s) can be determined. 

The spin Hamiltonian of the core-shell simulations is the sum of the respective Hamiltonian of the $L1_0$-phase core and of the $A1$-phase shell $\mathcal{H} = \mathcal{H}_{\mathrm{core}} + \mathcal{H}_{\mathrm{shell}}$ which, following the standard Heisenberg form, includes the exchange and anisotropy components without an external magnetic field $\mathbf{B}$ term. In general the exchange can be written in tensor form $-\left(\hat{\mathbf{S}}_i\right)^T \mathcal{\tilde{J}}_{ij} \hat{\mathbf{S}}_j$ where the  $\mathcal{\tilde{J}}_{ij}$ encapsulates anisotropic exchange and the Dzyaloshinskii-Moriya interaction. In the case of the $A1$ phase of FePt the exchange is isotropic while for the $L1_0$ phase the exchange is anisotropic and can be expressed~\cite{Mryasov_2005} as the sum of a diagonal tensor plus a two-ion anisotropy term $\mathcal{\tilde{J}}_{ij} = \mathcal{J}_{ij}+ 2\mathcal{K}_{\mathrm{2ion}}$. The Hamiltonian of the core and of the shell are given as follows:
\be
\begin{split}
    &\mathcal{H}_{\mathrm{core}} = -\frac{1}{2}\sum_{i,j\in tn} \hat{\mathbf{S}}_i^T \left(\mathcal{J}_{ij} + 2\mathcal{K}_{\mathrm{2ion}} \right) \hat{\mathbf{S}}_j - k^{L1_0}_{\mathrm{loc}} \sum_{i\in tn} (\hat{\mathbf{S}}_i^z)^2, \\
    &\mathcal{H}_{\mathrm{shell}} = -\frac{1}{2}\sum_{i,j\in tn} \hat{\mathbf{S}}_i^T \mathcal{J}_{ij} \hat{\mathbf{S}}_j - k^{A1}_{\mathrm{loc}} \sum_{i\in tn} (\hat{\mathbf{S}}_i^z)^2,
\end{split}
\label{Eq_Hamiltonian}
\ee
where $\hat{\mathbf{S}}_i$ and $\hat{\mathbf{S}}_j$ are spin unit vectors, $\mathcal{J}_{ij}$ the (isotropic) exchange energy tensor between pair $(i,j)$ within the truncated-neighbor range $tn$, and $k^{A1}_{\mathrm{loc}}$ and  $k^{L1_0}_{\mathrm{loc}}$ the local, in-plane single-site anisotropy of respective phase the numerical values of which in simulations are respectively set to 0~\cite{Schaaf} and -0.097~meV/atom~\cite{Mryasov_2005}. Unlike the exchange, the expression for anisotropy does not have to avoid double-counting, thus explaining the prefactor 2 of $\mathcal{K}_{\mathrm{2ion}}$ which cancels out the summation prefactor $1/2$. In FePt, the exchange interaction $\mathcal{J}_{ij}$ extends further than strictly nearest-neighbours. However, since the exchange interaction strength decreases rapidly with increasing distance between neighbouring atoms, a reasonably good model includes the exchange interactions truncated after the next-next-nearest neighbours. The calculation of the truncated $\mathcal{J}_{ij}$ and $\mathcal{K}_{\mathrm{2ion}}$ is described the Supplemental Material Sec.~S1~\cite{Supp} which yields a Curie temperature of around $700$~K for both ordered and disordered phases comparable with experiment~\cite{Hovorka}. Numerical values of exchange energy, anisotropies, and other simulation parameters are tabulated in Table \ref{Tab_SimulationParameters}. Simulations are repeated 10 times to compute statistical values. 
\begin{table}[tb!]
\centering
\begin{ruledtabular}
\fontsize{8}{6}
\begin{tabular}{l c c|l l l}
Parameter & Notation & Unit & $L1_0$-phase & A1-phase \\ 
\hline \hline
Atomistic spin moment & $\ensuremath{\mu_{\mathrm{s}}}$ & $\ensuremath{\mu_{\mathrm{B}}}$ & 3.23 & 3.23 \\
Local anisotropy & $k_{\mathrm{loc}}$ & meV/atom & - 0.097 & 0 \\
2-ion anisotropy & $k_{\mathrm{2ion}}$ & meV/atom & 1.427 & 0 \\
Total exchange~\cite{Supp}  & $J^0$ & J/link & $3\times10^{-21}$ & $3\times10^{-21}$ \\
\hline
CMC Equilibration steps &  &  & $2\times10^{5}$ & $2\times10^{5}$ \\ 
CMC Total step &  &  & $8\times10^{5}$ & $8\times10^{5}$ \\ 
\end{tabular}
\end{ruledtabular}
\caption[]{Coreshell simulation parameters}
\label{Tab_SimulationParameters} 
\end{table}
%

\begin{figure}[htbp!]
\centering
    \includegraphics[width=0.48\textwidth]{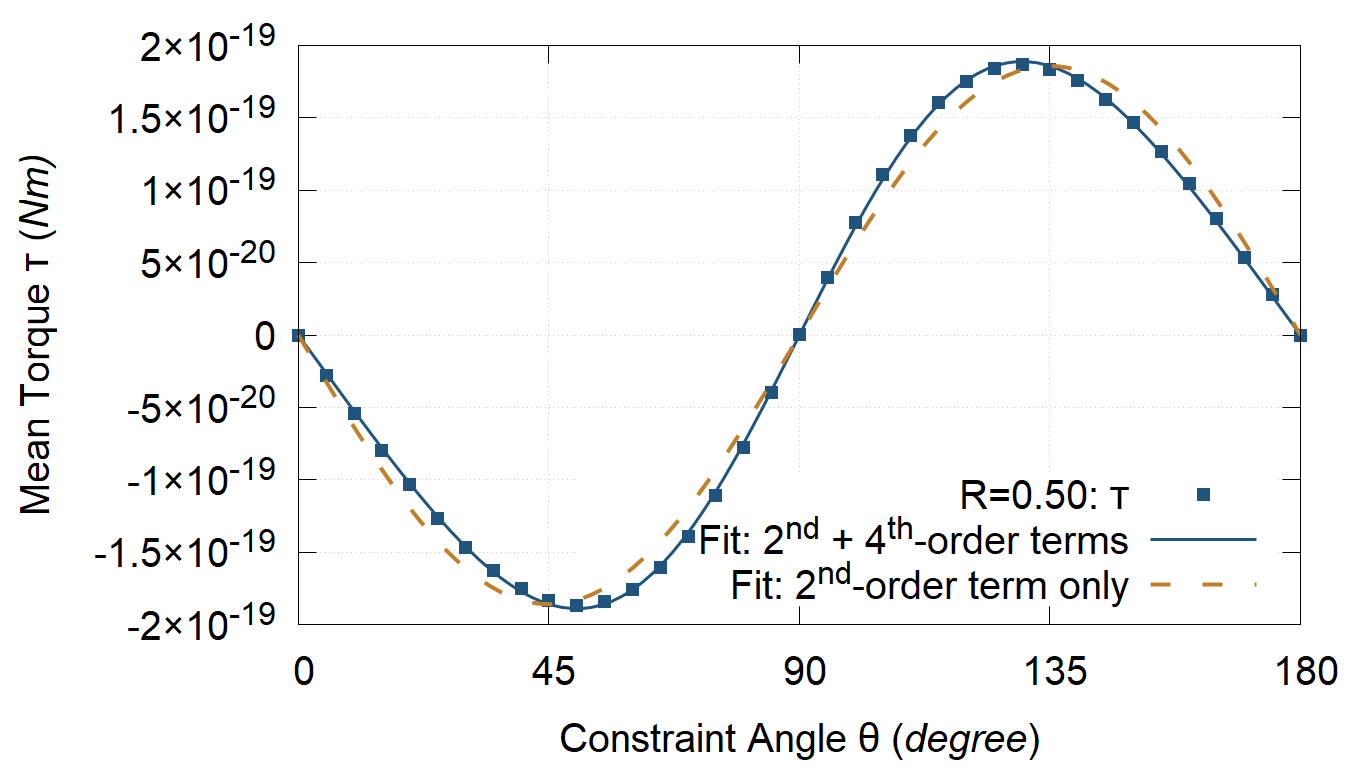}
    \includegraphics[width=0.48\textwidth]{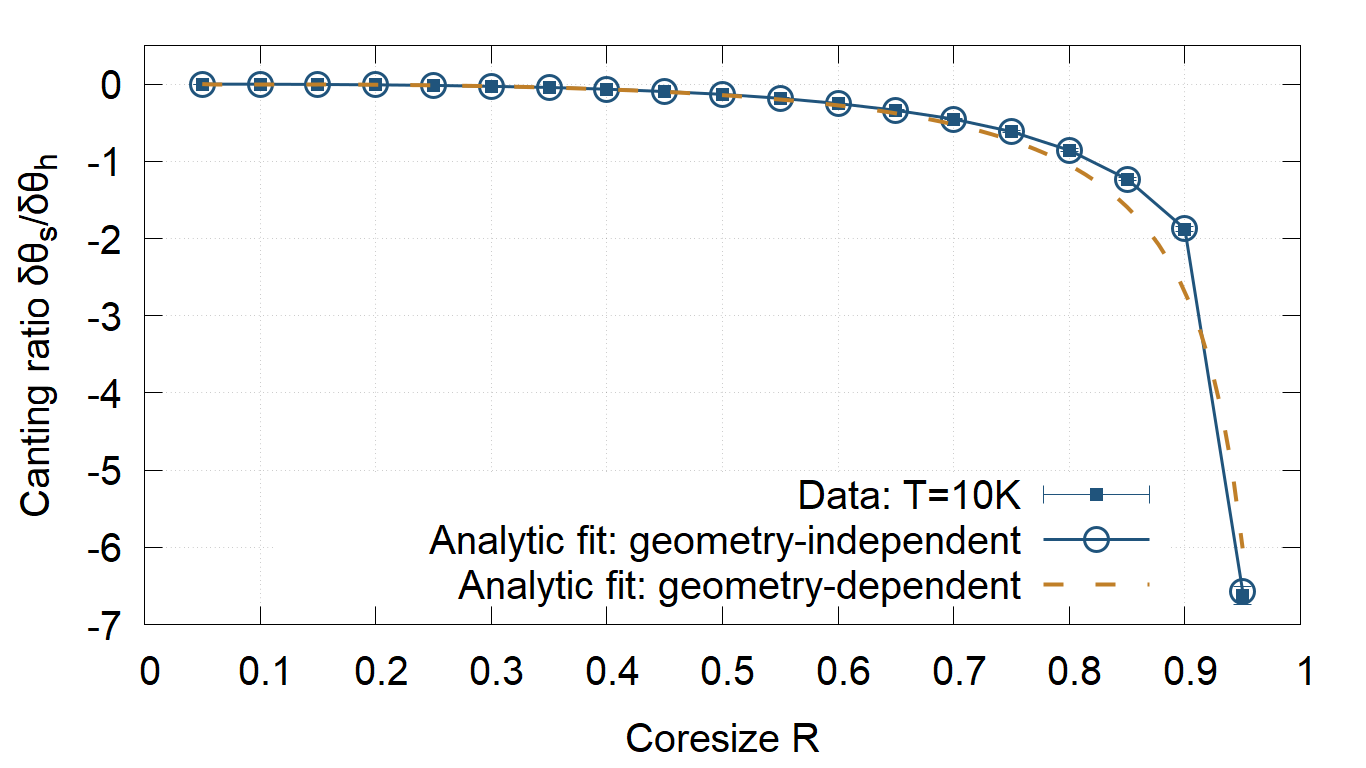}
\caption{(Upper) Fitting to the torque $\tau$ for a core-size $R=0.50$ at 10~K displays a clear deviation from simulation data (solid symbols) if including only a second-order anisotropy term (dashed line), but matches better if adding a fourth-order anisotropy term (solid line); (Lower) The canting of the core and shell magnetization with analytic fits.} 
\label{Fig_Canting}
\end{figure}

The magnetocrystalline anisotropy energy, if assumed to include only a second-order anisotropy term $E=K_1\sin^2(\theta)$, would imply a restoring torque $\tau(\theta) \propto \sin(2\theta)$. However, our simulation results - a sample shown in Fig.~\ref{Fig_Canting}(a) - demonstrate that a fit (dashed line) to the calculated torque  is noticeably skewed from the simulation data (solid symbols). In contrast, when a fourth-order anisotropy term is added i.e. $E(\theta) = K_1\sin^2(\theta) + K_2\sin^4(\theta)$, the new torque fit (solid line)  now matches the simulated data extremely well. The discernible skewing of the torque curve in comparison to a simple $\sin(2\theta)$ profile has been observed, without explanation, in a previous experimental study on FePt granular films~\cite{Saito_TorqueSkewed}. In our simulations, however, the skewed torque curves in Fig.~\ref{Fig_Canting}(a) are a clear indicator of the existence of 
a significant fourth-order anisotropy component in the core-shell grains.

\begin{figure}[htb!]
\centering
    \includegraphics[width=\linewidth]{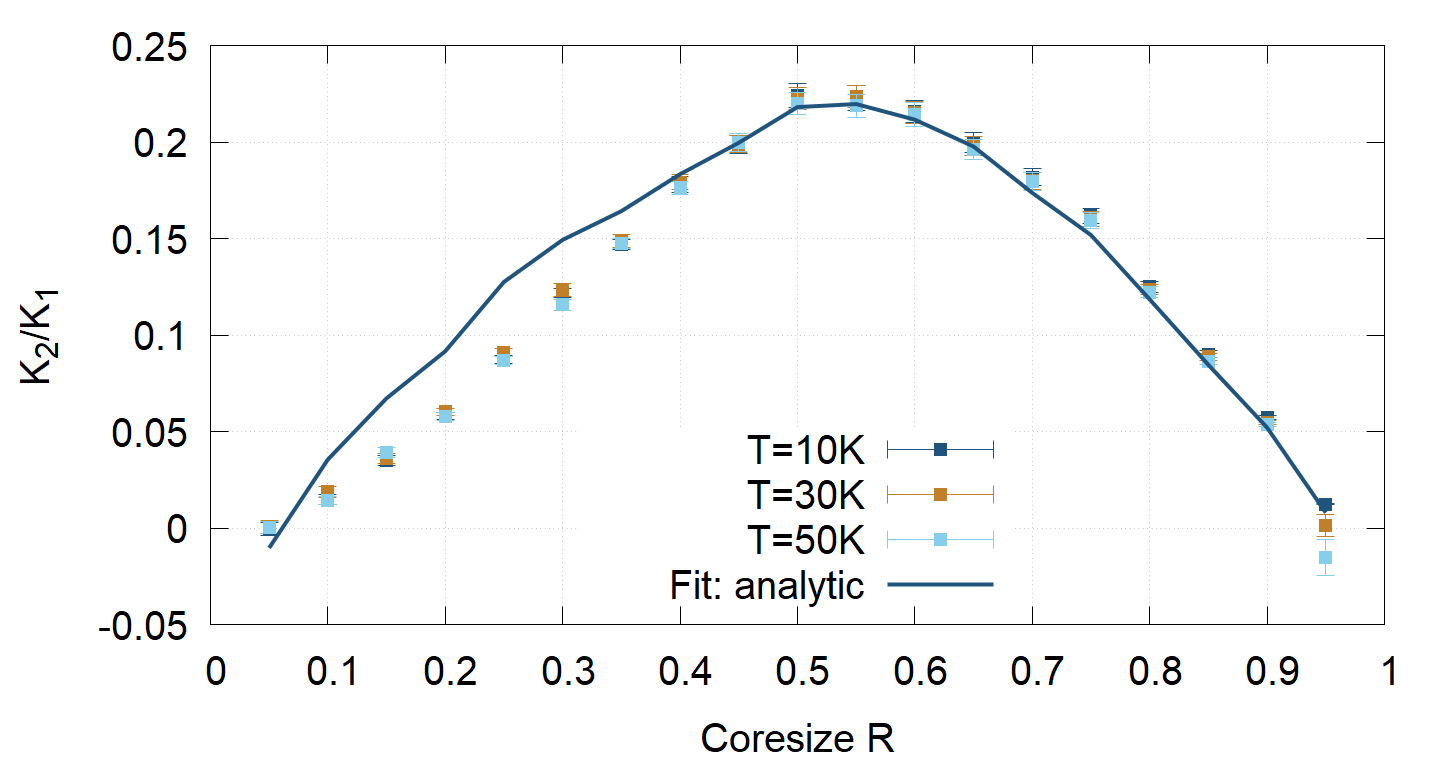}
    \caption{The dependence of $K_2/K_1$ on $R$. Symbols are the numerically determined values from atomistic simulations and the solid line is the predicted analytic value.}
    \label{Fig_KvsR}
\end{figure}

\begin{figure}[htb!]
\centering
    \includegraphics[width=\linewidth]{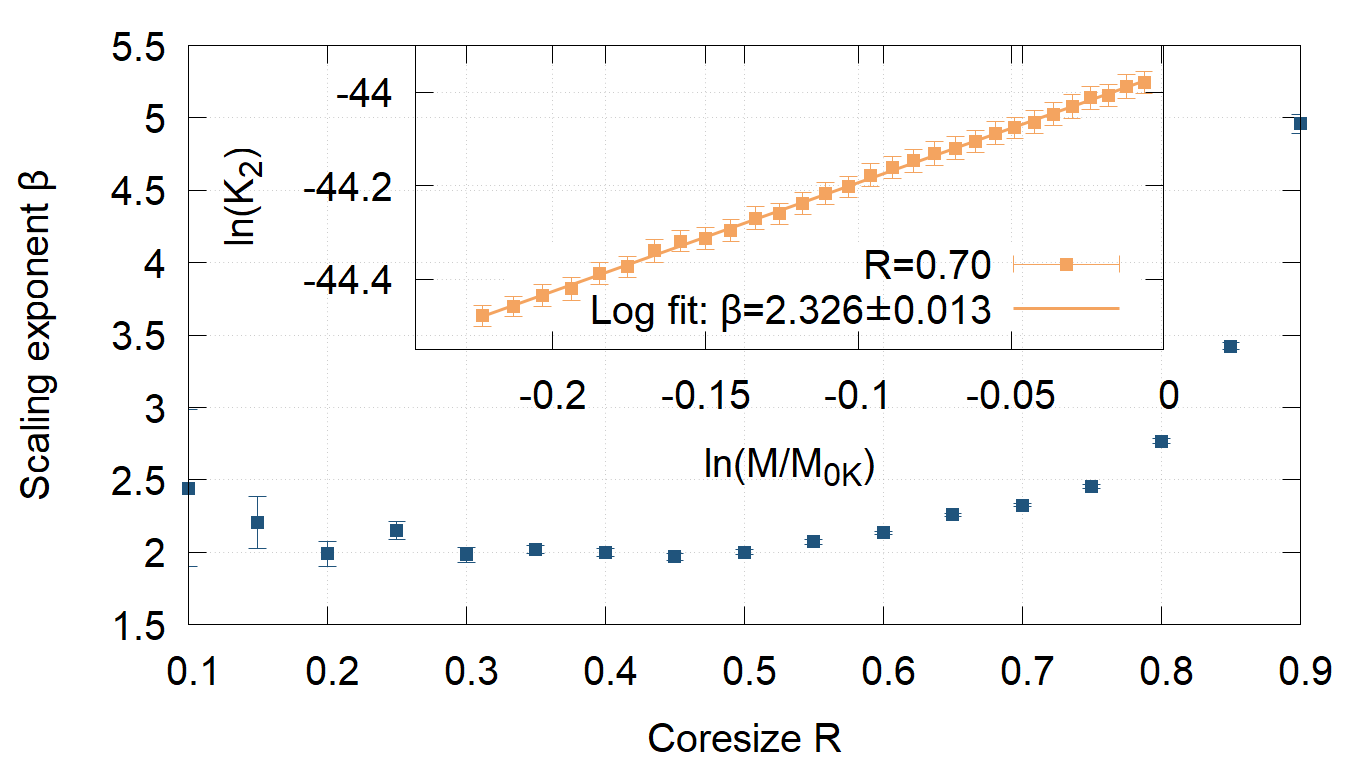}
    \caption{Variation of the scaling exponent $\beta$ of the fourth-order anisotropy constant $K_2$ as a function of the core size $R$ with an inset showing an example of $K_2$ scaling to M for R=0.70 with a scaling exponent $\beta=2.326  \pm 0.013$.}
    \label{Fig_Scaling}
\end{figure}

The magnitude of the temperature-dependent fourth-order anisotropy $K_2$, expressed via the $K_2/K_1$ ratio, is found to be dependent on the core size $R$ with a non-monotonic variation that has not been reported elsewhere. Low-temperature data in Fig.~\ref{Fig_KvsR} indicate that the magnitude of $K_2$ can be significant - exceeding 20\% of $K_1$ - if the proportions of the two phases are comparable, or insignificant - just a few percent of $K_1$ - if one phase dominates. This result is intriguing because it essentially recaptures conflicting observations in literature \cite{Inoue,Okamoto}. Additionally, the classical Callen-Callen power law~\cite{Callen1966TheLaw} predicts the scaling behavior $K_2 \propto (M/M_s)^{10}$, where $(M/M_s)$ is the magnetization normalized against the saturated magnetization $M_s$ at 0~K. However, our simulation results conclusively contradict this prediction. The inset of Fig.~\ref{Fig_Scaling} shows an example of $K_2$ scaling obtained from simulation for $R=0.70$ from which we find $\beta \approx 2.3$ only. Overall, the scaling exponent $\beta$ is found to be consistently lower than the Callen-Callen predicted value of 10. The variation of $\beta$ for $K_2$ as a function of $R$ - in Fig.~\ref{Fig_Scaling} - generally conforms to $2 \lessapprox \beta \lessapprox 3$ with exceptions seen in the two extreme cases $R=0.05$ and $R \ge 0.80$. 

%



To discuss these two key results, we propose a simple analytic model which we show can explain the origin and behaviors of the fourth-order anisotropy constant $K_2$ in not only our simulated FePt core-shell system but also a generic material with soft/hard magnetic interlayers. A full description of the analytic model is provided in the Supplemental Material Sec.~S2~\cite{Supp}, with the main points outlined as follows. The fundamental observation - illustrated in Fig.~\ref{Fig_Canting}(b) - is that there exists a canting between the core and shell magnetization which minimizes the total (interlayer exchange and anisotropy) energy of the system. Consider a general core-shell system in which the core is made of a hard-magnetic material having an out-of-plane uniaxial anisotropy $k_u$ (per atom) and the shell a soft-magnetic material with negligible uniaxial anisotropy and make the simplifying assumption of coherent magnetization in both core and shell. To minimize the system total magnetic energy $E$, the constrained angles of the magnetization $\theta_h$ and $\theta_s$ of the hard and soft-magnetic phase are allowed to deviate from the overall constraint angle $\theta$ by $\dth$ and $\dts$ respectively, which are sufficiently small to be treated as perturbations. Then $E$ can be expressed as: 
\be
    E = k_u N_c \sin^2(\theta+\delta \theta_h) - J N_{int}\cos(\delta \theta_h-\delta \theta_s),
\label{Eq_Energy}
\ee
where $N_c$ and $N_{int}$ are the number of spins in the core and the core/shell interface respectively, and $J$ the exchange integral. The first term in Eq.~\eqref{Eq_Energy} is an anisotropy term and the second an interlayer exchange term which describes the exchange coupling between the core spins and the shell spins. Since $\theta_h$ and $\theta_s$ are the averaged contributions of all spins in each respective phase, their proportional sum must result in the system $\theta$. Hence, Eq.~\eqref{Eq_Energy} must be minimized subject to:
\be
f\cos\theta_h + (1-f)\cos\theta_s - \cos\theta = 0,
\label{Eq_Constraint}
\ee
where $f = N_c/N_{tot}$ is the fractional volume of the hard phase with $N_{tot}$ being the total number of spins in the system. Note that for regular geometries, the geometry-independent variables $N_c$, $N_{int}$, and $N_{tot}$ can be replaced by the core volume $V_c$, the interface area $A_{cs}$, and the total volume $V_{tot}$ respectively. Substitute $\theta_h = \theta + \dth$ and $\theta_s = \theta + \dts$ into Eq.~\eqref{Eq_Constraint} and solve to first-order approximation:
\be
    \dts = -\frac{f}{1-f} \dth = -\frac{N_c/N_{tot}}{1-N_c/N_{tot}} \dth.
\label{Eq_AngleRatio}
\ee
Analytic fits following from Eq.~\eqref{Eq_AngleRatio} for both the geometry-dependent case $f=V_c/V_{tot}$ and the geometry-independence case $f=N_c/N_{tot}$ are shown in Fig.~\ref{Fig_Canting}(b). Because of the faceted shape of the simulated coreshell system, the geometry-dependent fit is seen to deviate from simulation data from $R \approx 0.75$ while the geometry-independent fit matches the entire data range extremely well. Minimizing Eq.~\eqref{Eq_Energy} subject to the constraint Eq.~\eqref{Eq_Constraint} leads to an expression of $E$ which explicitly includes both a second-order and a fourth-order anisotropy term the magnitude ratio of which is given by:
\be
    \frac{K_2}{K_1} \approx \frac{2k_uN_c}{JN_{int}}\left(1-\frac{N_c}{N_{tot}}\right)^2.
\label{Eq_K2K1}
\ee

To expand the model, consider cases similar to the simulated coreshell FePt in which the dominant part of the core uniaxial anisotropy $k_u$ comes from a 2-ion anisotropy. This necessitates two further considerations. First, the 2-ion anisotropy is lost on the core/shell interface because of the loss of Pt neighbors from the next-immediate atomistic layer of the shell. Second, when the core is small, the in-plane anisotropy becomes dominant because of vanishing 2-ion anisotropy at the core surface. Take the simulated coreshell FePt for example, the in-plane anisotropy is the sum of the local single-site anisotropies of the $L1_0$ and A1 phase, i.e. $k_{ip} = k^{L1_0}_{loc} + k^{A1}_{loc} = -0.097$~meV/atom, and manifests in the extreme case $R=0.05$ in Fig.~\ref{Fig_KvsR} from which $K_2/K_1$ is seen to become negative. Incorporating these two extra considerations into the first term of Eq.~\eqref{Eq_Energy} transforms its prefactor $k_u N_c$ to $\left[k_u \left(N_c - N_{int}\right) + k_{ip} N_c \right]$, which subsequently modifies the expression in Eq.~\eqref{Eq_K2K1} to:   
\be
    \frac{K_2}{K_1} \approx \frac{2k_uN_c}{JN_{int}}\left(1-\frac{N_{int}}{N_c}+\frac{k_{ip}}{k_u}\right)\left(1-\frac{N_c}{N_{tot}}\right)^2.
\label{Eq_K2K1General}
\ee

Fig.~\ref{Fig_KvsR} shows the variation of $K_2/K_1$ with $R$ at low temperatures in which analytic predictions (solid line) are compared with simulation data (symbols). Both analytic and numerical variations share a similar non-monotonic form with a peak attained at $R \approx 0.55$. The agreement for $R \geq 0.5$ is extremely good, while for $R < 0.5$ the $K_2/K_1$ values appear to be slightly over-estimated by the analytic model. Nonetheless the overall agreement is highly satisfactory which supports the hypothesis that the fourth-order anisotropy arises from the core/shell spin canting and the exchange energy contribution at the core/shell interface.

Furthermore, the scaling exponent $2 \lessapprox \beta \lessapprox 3$ of $K_2$ to $M/M_{s}$ as shown in Fig.~\ref{Fig_Scaling} can now be explained. Our analytic model finds that $K_2 \propto (k_u)^2/J$ - with a detailed derivation given in the Supplemental Material Sec.~S2~\cite{Supp}. It has been established for $L1_0$-FePt that $k_u \propto (M/M_s)^{2.1}$~\cite{Mryasov_2005,Skomski_K1_alloys,Skomski_K1_L10,Skomski_K1_PtCo,Staunton,Thiele_K1,Richard_Scaling} and $J \propto (M/M_s)^2$ via mean-field calculations~\cite{Atxitia2010,Atxitia2007}. The resulting scaling, therefore, reads $K_2 \propto (M/M_s)^{2.2}$ thus reproducing $\beta \approx 2.2$ in good agreement with simulation results. Finally, deviations from the analytic model for the two extreme cases $R=0.05$ and $R \ge 0.80$ can be explained from the previously mentioned inherent nature of the grain structure in the respective cases. In the $R=0.05$ case, the $L1_0$ core is so small that it consists of a single Fe atom which means the 2-ion anisotropy component of the $L1_0$ phase completely vanishes, leaving the core with just the negative in-plane single-site Fe anisotropy. This explains the negative ratio $K_2/K_1$ at $R=0.05$ as seen in Fig.~\ref{Fig_KvsR}. Meanwhile, for the $R \ge 0.80$ case, the $A1$ shell is so thin that it has exactly one or two atomistic layers, hence invalidating the fundamental premise of the analytic calculations which assumes interactions up to the next-next-nearest neighbors. Hence, $\beta$ of the $R \ge 0.80$ case was seen in Fig.~\ref{Fig_Scaling} to increase exponentially in the $ \beta \gtrapprox 3$ range, albeit still significantly lower than the Callen-Callen's predicted value of 10.   

%
In summary, we have presented a comprehensive study of higher-order anisotropy in a phase-coupled $L1_0$/$A1$-FePt coreshell system. A fourth-order anisotropy is found to exist due to a combination of the canting of the core and shell magnetization and the exchange coupling at the core/shell interface. This fourth-order anisotropy is demonstrated to exhibit a strong dependence on the system geometry and scale with $(M/M_s)^{2.2}$, which does not conform with the Callen-Callen power law. We formulate an analytic model to explain the origin and behaviors of this new fourth-order anisotropy from which a high level of agreement with numerical simulation is achieved. Overall, our findings provide substantial insights into a topic that has otherwise been lacking attention. Because anisotropy decides thermal stability of the writing medium, the significance of fourth-order anisotropy of $L1_0$-FePt can potentially translate to an issue of consideration for HAMR-related applications. Although investigated in the particular case of a $L1_0$/A1-FePt coreshell structure, the analytic model presented is valid for any combination of soft/hard materials. The phenomenon should therefore be observable in a wide variety of systems. 
%


\section{Acknowledgments}
%
The authors would like to thank Daniel Meilak for assistance with optimizing the simulated coreshell system. The financial support of the Advanced Storage Research Consortium (ASRC) is gratefully acknowledged. Sergiu Ruta acknowledges funding from the EPSRC TERASWITCH project (project ID EP/T027916/1). Sarah Jenkins acknowledges funding by the German Research Foundation (DFG) project No. 320163632. We are also grateful for computational support from the University of York High Performance Computing service, Viking, and the Research Computing team. 
%

\appendix

\bibliography{Reference}


\end{document}


\title{Higher-order magnetic anisotropy in soft-hard magnetic materials: Supplemental}
\author{Nguyen Thanh Binh \orcidlink{0000-0001-5869-6088}}
\affiliation{Department of Physics, University of York, York, YO10 5DD, UK}
\thanks{Corresponding author: btn500@york.ac.uk}
\author{Sarah~Jenkins\orcidlink{0000-0002-6469-9928}}
\affiliation{Department of Physics, University of York, York, YO10 5DD, UK}
\affiliation{TWIST Group, Institut f\"ur Physik, Johannes Gutenberg-Universit\"at, 55128 Mainz}
\affiliation{TWIST Group, Institut f\"ur Physik, University of Duisburg-Essen}
\author{Sergiu Ruta\orcidlink{0000-0001-8665-6817}}
\affiliation{Department of Physics, University of York, York, YO10 5DD, UK}
\affiliation{The Department of Engineering and Mathematics, Sheffield Hallam University, Sheffield, S1 1WB, UK}
\author{Richard~F.~L.~Evans\orcidlink{0000-0002-2378-8203}}
\affiliation{Department of Physics, University of York, York, YO10 5DD, UK}
\author{Roy~W.~Chantrell\orcidlink{0000-0001-5410-5615}}
\affiliation{Department of Physics, University of York, York, YO10 5DD, UK}
\maketitle

\section{S1 - The calculation of the truncated exchange interaction and 2-ion anisotropy}
\begin{figure}[ht!]
\centering
 \includegraphics[width=\linewidth]{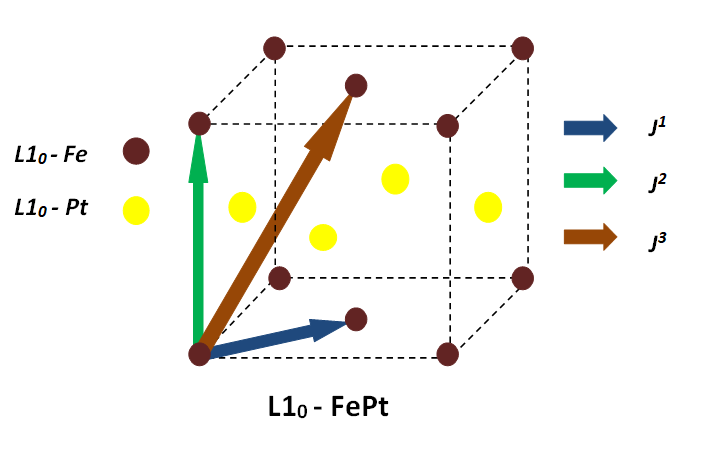}
   \caption{Visualisation of the truncation model in $L1_0$-FePt up to the next-next-nearest neighbors: Fe atoms are colored brown and Pt atoms yellow. Each $k^{th}-level$ nearest-neighbor has its own associated effective exchange interaction $J^k$.}
\label{Fig_ExchangedTruncated}
\end{figure}

This Section presents the calculation of the components of the truncated exchange interaction and the truncated 2-ion anisotropy implemented in the coreshell simulation in \vampire. We note that the separation of the 2-ion term in the $L1_0$ Hamiltonian leads to the same isotropic exchange value as for the $A1$ phase. Consequently we replace the tensor $\mathcal{J}_{ij}$ by a scalar form of the exchange. Here we present the derivation of the approximation used to truncate the long-ranged exchange interaction. The truncated exchange interaction strength can be expressed mathematically as:
\be
    J^0 = \sum_{k = 1,2,3}n_k J^k,
\label{Eq_ExchangeTruncated}
\ee
where $J^0$ is termed the "total" exchange interaction strength summing from all $k^{th}$-level nearest-neighbor components, $n_k$ the number of $k^{th}$-level nearest-neighbors, and $J^k$ the exchange interaction between $k^{th}$-level nearest-neighbors. Note the use of the term "$k^{th}$-level" prefixing "nearest-neighbors" is purely for simplicity, so that the $1^{st}$-level refers to the nearest neighbors, $2^{nd}$-level the next-nearest neighbors, and $3^{rd}$-level the next-next-nearest neighbors as illustrated in Fig.~\ref{Fig_ExchangedTruncated}.

Mryasov et al.'s model of the 2-ion anisotropy component of $L1_0$-FePt~\cite{Mryasov_2005} allows the removal of Pt atoms from the fct lattice structure, thus leaving only Fe-Fe interactions with modified properties. For the $A1$ phase, the random distribution of Fe and Pt atoms results in a highly chemically disordered structure with extremely low uniaxial anisotropy~\cite{Schaaf} in which the non-magnetic Fe-Pt interactions do not contribute to the magnetic properties of the material. Consequently, in \vampire the Pt atoms of both $A1$ and $L1_0$ phase of FePt are treated as non-magnetic and not accounted for in simulations. 

Because the exchange tensor $\mathcal{J}_{ij}$ applies to both the fct-lattice $L1_0$ and fcc-lattice $A1$ phases, the calculation of the exchange component $J^k$ will need to take into account different numbers of $k^{th}$-level nearest-neighbors $n_k$ in the lattice structure of each phase. The starting observation is that for an "artificial" bulk fcc-FePt configuration in which only nearest-neighbor interactions are included - denoted by the superscript $nn$, using an exchange energy strength $J^{\mathrm{nn}}_{\mathrm{fcc}} \approx 3\times10^{-21}$~J/link would yield a Curie temperature $T_c \approx 700$~K comparable with experiment~\cite{Hovorka}. In bulk fcc there are 12 nearest-neighbors in total, hence the total exchange interaction strength in bulk fcc is summed up as $J^0_{\mathrm{fcc}} =  12J^{\mathrm{nn}}_{\mathrm{fcc}}$. Since both the $A1$-FePt and $L1_0$-FePt are lattice-wise equivalent to the bulk fcc, the total exchange interaction strength of each phase can then be assigned similar numerical value $J^0_{A1} = J^0_{L1_0} = J^0_{\mathrm{fcc}} = 12J^{\mathrm{nn}}_{\mathrm{fcc}}$.    

In the $A1$ phase, an atom can have at maximum 12, 6, and 24 $1^{st}$-level, $2^{nd}$-level, and $3^{rd}$-level nearest neighbors respectively. However, statistically 50\% of total atoms in the $A1$ phase are the non-magnetic Pt so the averaged numbers of neighbors in the $A1$ phase need to be halved. Therefore Eq.~\eqref{Eq_ExchangeTruncated} applied for the $A1$ phase becomes:
\be
    J^0_{A1} = 6J^1 + 3J^2 + 12J^3 = 12J^{\mathrm{nn}}_{\mathrm{fcc}}.
\label{Eq_ExchangeTruncated_A1}
\ee
For the $L1_0$ phase with 4 Pt atoms in each middle layers being removed, there are 4, 6, and 8 nearest-neighbors of the $1^{st}$, $2^{nd}$, and $3^{rd}$-level respectively. Therefore the total exchange interaction strength in the $L1_0$ phase can be expressed as:   
\be
    J^0_{L1_{0}} = 4J^1 + 6J^2 + 8J^3 = 12J^{\mathrm{nn}}_{\mathrm{fcc}}.
\label{Eq_ExchangeTruncated_L10}
\ee
Another constraint is derived from Hinzke et al.~\cite{Hinzke} which found an estimated ratio of $3/4$ between out-of-plane total exchange to in-plane total exchange for the $L1_0$ phase. From Fig.~\ref{Fig_ExchangedTruncated}, the out-of-plane exchange comprises of all 8 $J^3$ components and 2 $J^2$ components along the [001] lattice direction, whilst the in-plane exchange includes all 4 $J^1$ components and the 4 remaining $J^2$ components along the [100] and [010] lattice direction. Therefore:
\be
    8J^3 + 2J^2 = \frac{3}{4}\left(4J^1 + 4J^2\right).
\label{Eq_ExchangeRatio_L10}
\ee
Solving Eq.~\eqref{Eq_ExchangeTruncated_A1}, Eq.~\eqref{Eq_ExchangeTruncated_L10}, and Eq.~\eqref{Eq_ExchangeRatio_L10} simultaneously yields the solution, to 7 decimals and being imported directly into \vampire simulations, as:
\be
\begin{split}
    &J^1_{xyz} = 0.7142857J^{\mathrm{nn}}_{\mathrm{fcc}} = 2.1428571\times 10^{-21}~\mathrm{J/link}, \\
    &J^2_{xyz} = J^{\mathrm{nn}}_{\mathrm{fcc}} = 3.0000000\times 10^{-21}~\mathrm{J/link}, \\
    &J^3_{xyz} = 0.3928571J^{\mathrm{nn}}_{\mathrm{fcc}} = 1.1785714\times 10^{-21}~\mathrm{J/link}.
\end{split}
\label{Eq_Exchange_Solution}
\ee

The calculation of the 2-ion anisotropy $\mathcal{K}_{\mathrm{2ion}}$ term of the $L1_0$ phase is done in a similar manner to the truncated exchange interaction. First, $\mathcal{K}_{\mathrm{2ion}}$ also takes the form of a 3-dimensional tensor which only acts on the z-component of the spins. Therefore all x and y components of $\mathcal{K}_{\mathrm{2ion}}$ vanish. Furthermore, $\mathcal{K}_{\mathrm{2ion}}$ is also truncated instead of being fully long-range, extending up to the $3^{rd}$-level of nearest-neighbors. The numerical values of the non-vanishing z-component of $\mathcal{K}_{\mathrm{2ion}}$ applicable for each $1^{st}$, $2^{nd}$, and $3^{rd}$-level nearest-neighbors are the solution to the following three simultaneous equations:
\be
\begin{split}
    &6k^1_z + 3k^2_z + 12k^3_z = k_{\mathrm{2ion}}, \\
    &4k^1_z + 6k^2_z + 8k^3_z = k_{\mathrm{2ion}}, \\
    &8k^3_z + 2k^2_z = \frac{3}{4}\left(4k^1_z + 4k^2_z\right),
\end{split}
\label{Eq_2ionAnisotropy}
\ee
where the 2-ion anisotropy constant is given by $k_{\mathrm{2ion}} = 12k^0 = 1.427$~meV/link~$= 2.2832\times10^{-22}$~J/link~\cite{Mryasov_2005}. The solution, similar to the truncated exchange solution, reads:
\be
\begin{split}
    &k^1_z = 0.7142857k^0 = 1.9026667\times 10^{-23}~\mathrm{J/atom}, \\
    &k^2_z = k^0 = 1.3590476\times 10^{-23}~\mathrm{J/atom}, \\
    &k^3_z = 0.3928571k^0 = 7.4747611\times 10^{-24}~\mathrm{J/atom}.
\end{split}
\label{Eq_2ionAnisotropy_Solution}
\ee

These treatments ensure that the simulated core-shell grains exhibit closely similar Curie temperatures for both the $L1_0$ core and the $A1$ shell consistent with experiment~\cite{SepehriAmin}. Additionally, the established scaling relation of the second-order anisotropy constant $K_1 \propto (M/M_s)^{2.1}$ for bulk $L1_0$-FePt~\cite{Mryasov_2005,Skomski_K1_alloys,Skomski_K1_L10,Skomski_K1_PtCo,Staunton,Thiele_K1,Richard_Scaling} is also successfully reproduced. 

\section{S2 - The analytic model}
In this Section, the full description of the analytic model outlined in the Main Text is presented from which a generalized expression for the fourth-order anisotropy constant is derived. This fourth-order anisotropy is shown to arise from an interface between two magnetic materials and to depend strongly on the system geometry. 
%

\subsection{Geometry-dependent formulation}
The fundamental observation is that there exists a canting between the core and shell magnetization which minimizes the total (interlayer exchange and anisotropy) energy of the system. Consider a general coreshell system in which the core is made of a hard-magnetic material having an out-of-plane uniaxial anisotropy constant $K_u$ (per volume) and the shell a soft-magnetic material with negligible uniaxial anisotropy and make the simplifying assumption of coherent magnetization in both core and shell. To minimise the system total (interlayer exchange and anisotropy) magnetic energy, the constrained angles of the magnetization $\theta_h$ and $\theta_s$ of the hard and soft-magnetic phase are allowed to deviate from the overall constraint angle $\theta$ by $\dth$ and $\dts$ respectively, which are sufficiently small to be treated as perturbations. The system total magnetic energy $E$ can be expressed as: 
\be
    E = K_uV_c\sin^2(\theta+\delta \theta_h) - \frac{J A_{cs}}{a^2}\cos(\delta \theta_h-\delta \theta_s),
\label{Eq_Energy}
\ee
where $V_c$ is the volume of the hard-phase core, $A_{cs}$ the area of the core/shell interface, $a$ the lattice spacing, and $J$ the exchange integral (per link). For simplicity, define the two coefficients:  
\be
\begin{split}
    &A= K_uV_c, \\
    &B= \frac{J A_{cs}}{a^2},
\label{Eq_Coefficients}
\end{split}
\ee
from which $B \gg A$ since typically $J \gg K_u$. The first term in Eq.~\eqref{Eq_Energy} is an anisotropy term and the second an interlayer exchange term which describes the exchange coupling between the core spins and the shell spins. Since $\theta_h$ and $\theta_s$ are the averaged contributions of all spins in each respective phase, their proportional sum results in the system $\theta$. Therefore, Eq.~\eqref{Eq_Energy} must be minimised subject to:
\be
f\cos\theta_h + (1-f)\cos\theta_s - \cos\theta = 0,
\label{Eq_Constraint}
\ee
where $f$ is the fractional volume of the hard phase core. Substitute $\theta_h = \theta + \dth$ and $\theta_s = \theta + \dts$ into Eq.~\eqref{Eq_Constraint} and solve to first-order approximation:
\be
    \dts = -\frac{f}{1-f} \dth.
\ee
Denote $F=f/(1-f)$. Then minimise Eq.~\eqref{Eq_Energy} subject to the constraint Eq.~\eqref{Eq_Constraint} by differentiating $E$ w.r.t. $\dth$, noting that $\dth - \dts  = \dth+F\dth =\dth(1+F)$:  
\be 
    A\sin\left[2\left(\theta + \delta \theta_h\right)\right] + B (1+F) \sin\left[\delta \theta_h\left(1 + F\right)\right] = 0.
\ee
The solution of which, up to the first-order of $\dth$, reads:
\be
    \dth  = -\frac{\sin \theta \cos \theta}{\frac{B}{2A}(1+F)^2}.  
\ee
Expand the trigonometric part of the anisotropy term in Eq.~\eqref{Eq_Energy} to first-order approximation, substituting in $\dth$ expression:
\be
\begin{split}
    \sin^2(\theta + \dth ) &= (\sin\theta\cos\dth + \cos\theta\sin\dth)^2 \\
    &= \sin^2\theta\cos^2\dth + 2\sin\theta\cos\dth\cos\theta\sin\dth \\
    &+ \cos^2\theta\sin^2\dth \\
    &\approx \sin^2\theta + \left(2\sin\theta\cos\theta\right)\dth  \\
    &= \sin^2\theta - \frac{2\sin^2\theta\cos^2 \theta}{\frac{B}{2A}(1+F)^2}\\
    &= \sin^2\theta - \frac{\sin^2\theta}{\frac{B}{4A}(1+F)^2} + \frac{\sin^4\theta}{\frac{B}{4A}(1+F)^2}.
\end{split}
\label{Eq_AnisotropyTerm}
\ee 
Next, expand the trigonometric part of the exchange term in Eq.~\eqref{Eq_Energy} to first-order approximation, noting that since $B \gg A$ the cosine argument is small: 
\be
\begin{split}
    \cos (\dth -\dts) &= \cos\left(\dth(1+F)\right) =  \cos\left(\frac{-\sin\theta\cos\theta}{\frac{B}{2A}(1+F)}\right)   \\
    &\approx 1 - \frac{\sin^2\theta\cos^2\theta}{2\left[\frac{B}{2A}(1+F)\right]^2} \approx - \frac{\sin^2\theta\cos^2\theta}{2\left[\frac{B}{2A}(1+F)\right]^2} \\
    &= -\frac{\sin^2\theta}{\frac{B^2}{2A^2}(1+F)^2} +\frac{\sin^4\theta }{\frac{B^2}{2A^2}(1+F)^2}.
\end{split}
\label{Eq_ExchangeTerm}
\ee
Note that the constant of 1 has been dropped. Substituting Eq.~\eqref{Eq_AnisotropyTerm} and Eq.~\eqref{Eq_ExchangeTerm} into Eq.~\eqref{Eq_Energy} gives:
\be
\begin{split}
    E &= A\left[\sin^2\theta - \frac{\sin^2\theta}{\frac{B}{4A}\left(1+F\right)^2} + \frac{\sin^4\theta}{\frac{B}{4A}(1+F)^2}\right] \\
    &- B\left[-\frac{\sin^2\theta}{\frac{B^2}{2A^2}(1+F)^2} + \frac{\sin^4\theta}{\frac{B^2}{2A^2}(1+F)^2}\right] \\
    &= \left[A - \frac{4A^2}{B(1+F)^2} + \frac{2A^2}{B(1+F)^2}\right]\sin^2\theta \\
    &+ \left[\frac{4A^2}{B(1+F)^2} - \frac{2A^2}{B(1+F)^2} \right]\sin^4\theta \\
    &= \left[A - \frac{2A^2}{B(1+F)^2}\right]\sin^2\theta + \frac{2A^2}{B(1+F)^2}\sin^4\theta \\
    &= \left[A - \frac{2A^2}{B}(1-f)^2\right]\sin^2\theta + \left[\frac{2A^2}{B}(1-f)^2\right]\sin^4\theta.
\end{split}
\label{Eq_Energy_Explicit}
\ee
Evidently, the core/shell interaction has mathematically introduced a fourth-order anisotropy term with a coefficient:
\be
    K_2 V_c = \frac{2A^2}{B}(1-f)^2.
\label{Eq_K2}
\ee
Also, the coefficient of the second-order anisotropy term, instead of just $A = K_uV_c$, is subsequently modified to become:
\be
    K_1 V_c = A - \frac{2A^2}{B}(1-f)^2.
\label{Eq_K1}
\ee
The ratio of which, noting that $B \gg A$, therefore is given by:
\be
\begin{split}
    \frac{K_2}{K_1} &= \frac{\frac{2A^2}{B}(1-f)^2}{A - \frac{2A^2}{B}(1-f)^2}
    = \frac{2A(1-f)^2}{B-2A(1-f)^2} \\ &\approx \frac{2A}{B}(1-f)^2 
    = \frac{2K_u V_c}{J A_{cs}/a^2}(1-f)^2.
\end{split}
\label{Eq_Ratio}
\ee

We then demonstrate, using Eq.~\eqref{Eq_Ratio}, the calculation of the maximum location of the fourth-order anisotropy constant $K_2$ for specific cases of highly symmetric coreshell geometries. For each case, $R$ remains to be the $L1_0$-phase coresize ratio which indicates the relative thickness of the $L1_0$-phase to the entire coreshell grain.
\begin{itemize}
    \item Planar bilayer with fixed surface sizes and variable thickness $L$:
\be
\begin{split}
    \frac{K_2}{K_1} &= \frac{2K_u V_c}{JA_{cs}/a^2}(1-R)^2 
    = \frac{2K_u\left[ (LR)A_{cs}\right]}{JA_{cs}/a^2}(1-R)^2 \\
    &= \frac{2K_u L}{J/a^2}R(1-R)^2.
\end{split}
\ee
    Thus $K_2/K_1$ attains the maximum at $R=1/3 \approx 0.333$.
    \item Cylinder with fixed height and variable base radius $L$:
\be
\begin{split}
    \frac{K_2}{K_1} &= \frac{2K_uV_c}{JA_{cs}/a^2}(1-R^2)^2
    = \frac{2K_u\left[\pi (LR)^2h\right]}{J\left[2\pi (LR)h\right]/a^2}(1-R^2)^2 \\
    &= \frac{K_uL}{J/a^2}R(1-R^2)^2.
\end{split}
\ee
    Thus $K_2/K_1$ attains the maximum at $R=1/\sqrt5 \approx 0.447$.
    \item Sphere with variable radius $L$:
\be
\begin{split}
    \frac{K_2}{K_1} &= \frac{2K_uV_c}{JA_{cs}/a^2}(1-R^3)^2 = \frac{2K_u\left[\frac{4}{3}\pi (LR)^3\right]}{J\left[4\pi (LR)^2\right]/a^2}(1-R^3)^2 \\
    &= \frac{\frac{2}{3}K_uL}{J/a^2}R(1-R^3)^2.
\end{split}
\ee
    Thus $K_2/K_1$ attains the maximum at $R=1/\sqrt[3]7 \approx 0.523$ which is reasonably close to the maximum location of the simulated coreshell grains at $R \approx 0.5$ seen before in Fig.~3 of the Main Text. 
\end{itemize}
%

%
Eq.~\eqref{Eq_Ratio} applies well for a system with only a uniaxial anisotropy constant $K_1$. To expand the model, consider cases similar to the simulated coreshell FePt in which the dominant part of the core uniaxial anisotropy $K_u$ comes from a 2-ion anisotropy. This necessitates two further considerations. First, the 2-ion anisotropy is lost on the core/shell interface because of the loss of Pt neighbors from the disordered $A1$ shell. A reasonable assumption can be made here that only the atomistic surface layer of the core will lose the 2-ion anisotropy from the next-immediate atomistic surface layer of the shell. Second, when the core is small, the in-plane anisotropy becomes dominant because of vanishing 2-ion anisotropy at the core surface. For example, consider the simulated coreshell FePt in which the in-plane anisotropy $K_{ip}$ (per volume) is the sum of the local single-site anisotropies of the $L1_0$ and A1 phase i.e. $K_{ip} = K^{L1_0}_{loc} + K^{A1}_{loc}$ and manifests in the extreme case $R=0.05$ which from Fig.~3 of the Main Text $K_2/K_1$ is seen to become negative. Incorporating these two extra considerations into the first term of Eq.~\eqref{Eq_Energy} transforms its prefactor $K_u V_c$ to $\left[K_u(V_c - aA_{cs}) + K_{ip}V_c\right]$, which following the same procedure subsequently modifies Eq.~\eqref{Eq_Ratio} to:   
\be
\begin{split}
    \frac{K_2}{K_1} &\approx \frac{\left[2K_u(V_c - aA_{cs}) + K_{ip}V_c\right]}{J A_{cs}/a^2}\left(1-f\right)^2 \\ 
    &= \frac{2K_u V_c}{J A_{cs}/a^2}\left( 1-\frac{aA_{cs}}{V_c}+\frac{K_{ip}}{K_u} \right)\left(1-f\right)^2.
\end{split}
\label{Eq_Ratio_Inplane}
\ee
In comparison to Eq.~\eqref{Eq_Ratio}, the addition of the in-plane anisotropy and the reduction of 2-ion anisotropy due to surface neighbor loss has induced an extra "conversion" factor given by the first bracket in  Eq.~\eqref{Eq_Ratio_Inplane}. 

\subsection{General geometry-independent formulation}
The faceted, elongated coreshell structure investigated numerically in this paper cannot be accurately approximated as one of the high-symmetry cases given earlier. Without reference to a specific geometry, the problem needs to be reformulated in terms of the number of spins/atoms in specific locations as follows. First, the energy given in Eq.~\eqref{Eq_Energy} can be rewritten as:
\be
    E = k_u N_c  \sin^2(\theta+\delta \theta_h) - J N_{int}\cos(\delta \theta_h-\delta \theta_s),
\ee
where the geometry-dependant variables $V_c$ and $A_{cs}/a^2$ have been replaced by their respective geometry-independent counterparts the number of spins in the core $N_c$ and the number of spins in the core/shell interface $N_{int}$. Additionally, the uniaxial anisotropy constant $K_u$ (per volume) is replaced by the equivalent $k_u$ (per atom). The two coefficients $A$ and $B$ from Eq.~\eqref{Eq_Coefficients} now reads $A = k_uN_c$ and $B = J N_{int}$, and likewise $f=N_c/N_{tot}$ with $N_{tot}$ being the total number of spins in the system. Substituting these into Eq.~\eqref{Eq_K1}, Eq.~\eqref{Eq_K2}, and Eq.~\eqref{Eq_Ratio} leads to new expressions:
\be
\begin{split}
    &K_1 N_c = k_uN_c - \frac{2(k_uN_c)^2}{JN_{int}}\left(1-\frac{N_c}{N_{tot}}\right)^2, \\
    &K_2 N_c = \frac{2(k_uN_c)^2}{JN_{int}}\left(1-\frac{N_c}{N_{tot}}\right)^2, \\
    &\frac{K_2}{K_1} \approx \frac{2k_uN_c}{JN_{int}}\left(1-\frac{N_c}{N_{tot}}\right)^2.
\end{split}
\label{Eq_Solution_General}
\ee
In order to incorporate the in-plane anisotropy (per atom) $k_{ip}$, substitute $A_{cs}=a^2 N_{int}$ and $V_c=a^3 N_c$ into Eq.~\eqref{Eq_Ratio_Inplane} to obtain:
\be
    \frac{K_2}{K_1} \approx \frac{2k_uN_c}{JN_{int}}\left(1-\frac{N_{int}}{N_c}+\frac{k_{ip}}{k_u}\right)\left(1-\frac{N_c}{N_{tot}}\right)^2.
\label{Eq_Ratio_Inplane_General}
\ee

A further generalization is to note that the expression in Eq.~\eqref{Eq_Ratio_Inplane_General} still retains the assumption that only one atomistic surface layer of the core loses 2-ion anisotropy from the next-immediate surface layer of the shell. This assumption, for cases similar to the simulated coreshell FePt with truncated exchange and 2-ion anisotropy, is only an approximation. Consider again the 2-ion anisotropy in the simulated coreshell FePt which, in principle, it is quite long-ranged. However, using the truncation model, a simplified representation of this 2-ion anisotropy can be made which involves up to the next-next-nearest neighbors of Fe spins. Since the disordered $A1$-phase shell has negligible anisotropy, the anisotropy energy part $E_{\mathrm{ani}}$ - i.e. the first term - of Eq.~\eqref{Eq_Energy} can be written as the sum of contributions over all spins of the ordered $L1_0$-phase core including both the 2-ion and the in-plane single-ion $k_{ip}$ components:
\be
    E_{\mathrm{ani}} = \sum_{i=1}^{N_c} \sum_{j=1}^3n^j_{i}k^j_{i} + k_{ip}N_c,
\label{Eq_AnisotropyEnergy_Exact}
\ee
where the 2-ion anisotropy component is described by the summation. The superscript $j$ indicates the level of nearest-neighbors within the truncation range; $n^j_{i}$ and $k^j_{i}$ are respectively the numbers of $j^{th}$-level nearest-neighbors and their corresponding anisotropy contribution as imported from Eq.~\eqref{Eq_2ionAnisotropy_Solution}. A coresize-dependent effective uniaxial anisotropy (per atom) $k^{\mathrm{eff}}_u(R)$ which incorporates the in-plane component $k_{ip}$ can be then defined as: 
\be
    k_u^{\mathrm{eff}}(R) = \frac{1}{N_c} \sum_{i=1}^{N_c} \sum_{j=1}^3n^j_{i}k^j_i + k_{ip}, 
\label{Eq_Anisotropy_Effective}
\ee
which allows the $E_{\mathrm{ani}}$ in Eq.~\eqref{Eq_AnisotropyEnergy_Exact} to be expressed simply by $E_{\mathrm{ani}}=k_u^{\mathrm{eff}} N_c$. Eq.~\eqref{Eq_AnisotropyEnergy_Exact} has therefore described the exact expression for the anisotropy energy rather than the surface-monolayer approximation which has been previously used.

In bulk $L1_0$-FePt, for all $i$, $n^j_{i}=4,6,8$ for $j=1,2,3$ respectively. However, in a general finite-size coreshell system this is no longer the case because of the loss of coordination at the surface. Hence in analytic calculations, $n^j_{i}$ has to be determined numerically for the specific finite-size coreshell system in investigation. The exact expressions for $K_1$, $K_2$, and $K_2/K_1$ can be obtained by substituting $k^{\mathrm{eff}}_u$ in Eq.~\eqref{Eq_Anisotropy_Effective} to Eq.~\eqref{Eq_Solution_General}: 
\be
\begin{split}
    &K_1 N_c = k^{\mathrm{eff}}_u N_c - \frac{2(k^{\mathrm{eff}}_u N_c)^2}{JN_{int}}\left(1-\frac{N_c}{N_{tot}}\right)^2, \\
    &K_2 N_c= \frac{(k^{\mathrm{eff}}_u N_c)^2}{JN_{int}}\left(1-\frac{N_c}{N_{tot}}\right)^2, \\
    &\frac{K_2}{K_1} = \frac{(k^{\mathrm{eff}}_u N_c)}{JN_{int}}\left(1-\frac{N_c}{N_{tot}}\right)^2.
\end{split}
\label{Eq_Solution_General_Effective}
\ee

\begin{figure}[ht!]
  \centering
  \includegraphics[width = \linewidth]{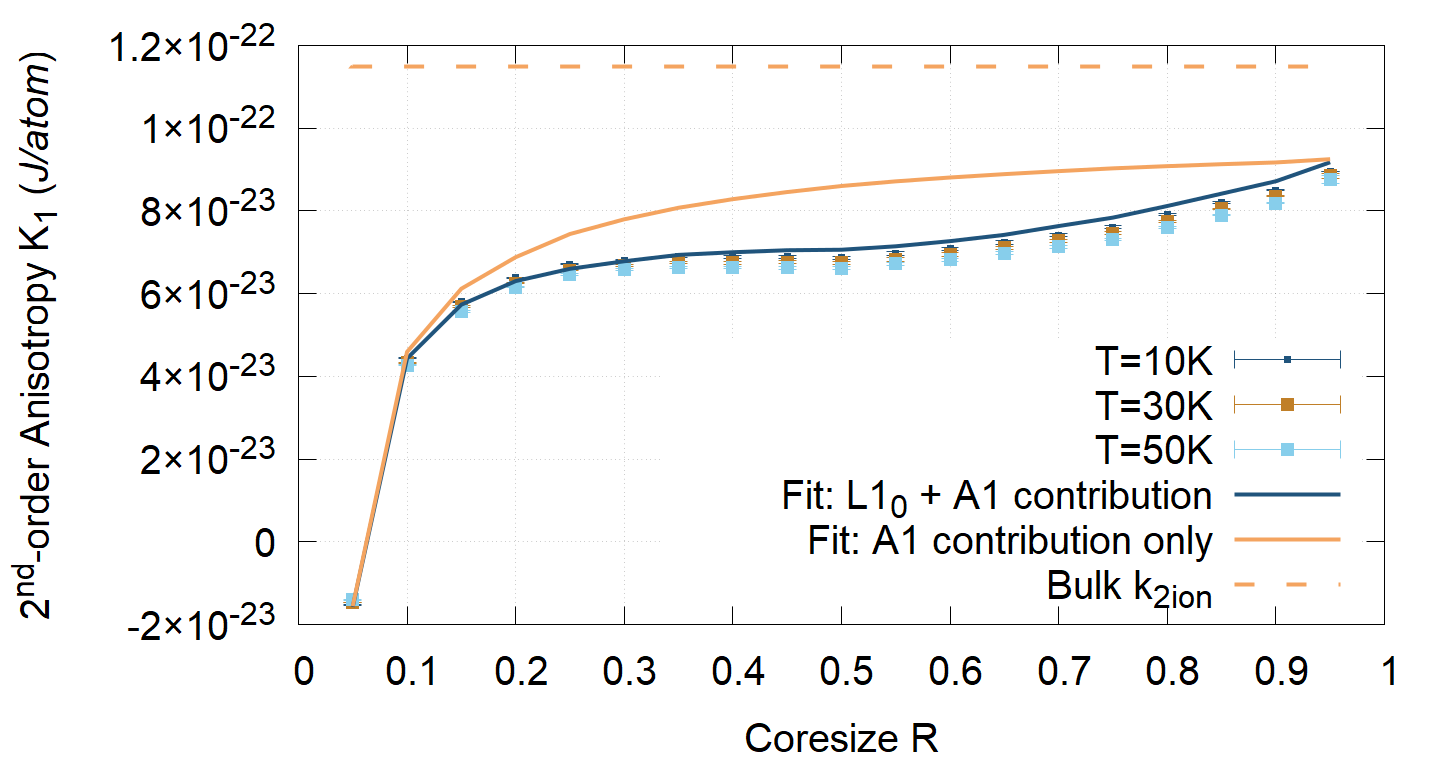}
  \caption[]{The variation of the second-order anisotropy constant $K_1$ (per atom) with the coresize $R$: symbols are atomistic simulation data; the solid blue line is the analytic fit including contributions from both the $L1_0$ and $A1$ phases; the solid brown line is the analytic fit excluding the contribution of the $A1$ phase; and the dashed brown line is the known value for bulk $k_{\mathrm{2ion}}$ in literature.}
\label{Fig_K1vR_PerAtom}
\end{figure}

In Fig.~\ref{Fig_K1vR_PerAtom} the variation of the $K_1$ (per atom) against $R$ at low temperatures is illustrated with two analytic fits for comparison: first a fit derived from Eq.~\eqref{Eq_Solution_General_Effective} which includes both the $L1_0$-phase and the $A1$-phase contributions, and second a fit $k_u^{\mathrm{eff}}(R)$ from Eq.~\eqref{Eq_Anisotropy_Effective} that excludes the $A1$-phase contribution. The first analytic fit is seen to match simulation data extremely well, which demonstrates that $K_1$ variation is a combination of both the 2-ion and in-plane single-ion anisotropies as well as the interlayer exchange induced by the presence of the $A1$ phase. For small $R$, the anisotropy is dominated by the in-plane anisotropy and increases from a small negative value as the larger 2-ion term begins to dominate. Rather than reaching an asymptotic value, $K_1$ increases monotonically as the loss of anisotropy at the surface decreases relatively in importance. Furthermore, the second analytic fit shows that without the contribution of the $A1$ phase which gives rise to the fourth-order $K_2$, $K_1$ will increase asymptotically to converge to the well-known value of the bulk 2-ion anisotropy $k_{\mathrm{2ion}}$ of $L1_0$-FePt. Overall, the agreement is excellent, thus validating the formulation of the effective uniaxial anisotropy $k^{\mathrm{eff}}_u(R)$ term. 
%


\bibliography{Reference}